\documentclass{article}
\usepackage[utf8]{inputenc}

\title{OncoPetNet: A Deep Learning based AI system for mitotic figure counting on H\&E stained whole slide digital images in a large veterinary diagnostic lab setting}
\author{Michael Fitzke$^1$, Derick Whitley$^2$, Wilson Yau$^2$,  \\ Fernando Rodrigues Jr$^1$, Vladimir Fadeev$^1$, Cindy Bacmeister$^2$\\ 
Chris Carter$^1$, Jeffrey Edwards$^2$,  Matthew P. Lungren$^3$, Mark Parkinson$^1$}
\date{%
    $^1$Mars Digital Technologies\\%
    $^2$Antech Diagnostics\\%
    $^3$Stanford University\\[2ex]%
}

\usepackage{fancyhdr}
\usepackage[numbers]{natbib}
\usepackage{graphicx}
\usepackage{amsmath}
\usepackage{amssymb}
\usepackage{hyperref}
\usepackage{lscape} 
\usepackage{graphicx}
\usepackage{geometry}
\usepackage{mathrsfs}
\usepackage{subfig}
\usepackage{float}
\graphicspath{ {./images/} }

\begin{document}

\maketitle

\begin{abstract}

\par 
\textbf{Background:} Histopathology is an important modality for the diagnosis and management of many diseases in modern healthcare, and plays a particularly critical role in cancer care. Pathology samples can be large and require multi-site sampling, leading to upwards of 20 slides for a single tumor, and the human-expert tasks of site selection and and quantitative assessment of mitotic figures are time consuming and subjective, particularly in highly mitotically active tumors.Automating these tasks in the setting of a digital pathology service presents significant opportunities to improve workflow efficiency and augment human experts in practice, yet major technical challenges remain that limit achieving use of these systems at scale in clinical workflows.
\par
\textbf{Approach:} Multiple state-of-the-art deep learning techniques for whole slide histopathology image classification and mitotic figure detection were used in the development of OncoPetNet. Additionally, model-free approaches were used to increase speed and accuracy in real-time deployment. The robust and scalable inference engine leverages Pytorch's performance optimizations as well as specifically developed speed up techniques in inference.
\par
\textbf{Results:} The proposed deep-learning system, OncoPetNet, demonstrated significantly improved mitotic counting performance for 41 cancer cases across 14 cancer types compared to human expert baselines. Further, in 21.9\% of cases use of the OncoPetNet led to change in tumor grading compared to human expert evaluation. In deployment, an effective 0.27 min/slide inference was achieved in a high throughput multi-site veterinary diagnostic pathology service across 2 centers processing 3,323 digital whole slide images daily.  
\par
\textbf{Conclusion:} This work represents the first successful automated deployment of deep learning systems for real-time expert-level performance on important histopathology tasks at scale in a high volume clinical practice. The resulting impact outlines important considerations for model development, deployment, clinical decision making, and informs best practices for implementation of deep learning systems in digital histopathology practices.

\end{abstract}

\section{Introduction}

Counting mitotic figures (MF) in hematoxylin and eosin-stained histologic slides is an important and time-intensive component of the diagnostic pathologist's evaluation of cancer.(\cite{donovan2020mitotic})  The mitotic count (MC) is a uniquely quantitative metric and one of the most valuable prognostic factors identified in the diagnostic workup of a pet with cancer. For example, in veterinary medicine, MC plays a principle role in many established grading schemes (\cite{donovan2020mitotic}) including canine mast cell tumors(\cite{kiupel2011proposal}), canine soft tissue sarcomas(\cite{dennis2011prognostic}, \cite{mcsporran2009histologic}), canine melanocytic tumors\cite{spangler2006histologic}), canine and feline mammary tumors (\cite{pena2013prognostic},\cite{mills2015prognostic}), as well as serving as an independent prognostic factor in many tumor types.  Recent studies have highlighted the subjective nature of the mitotic count, with many studies indicating significant inter-observer variation independent of region selection.(\cite{aubreville2020deep}, \cite{bertram2020computerized}).  In addition to challenges with inter-observer variation, the mitotic count is time consuming and an inefficient manual task.  

Recent advancements in deep learning and the availability of large datasets have enabled algorithms to match the performance of medical professionals in a wide variety of medical imaging tasks, including diabetic retinopathy detection (\cite{gulshan2016development}), skin cancer classification (\cite{esteva2017dermatologist}), and lymph node metastases detection (\cite{bejnordi2017diagnostic}). Only recently has the increasing trend toward digitization of whole slide pathology images (WSI) paved the way for fully automated analysis of histological slides, even approaching the accuracy of pathologists for certain well-defined tasks (\cite{bejnordi2017diagnostic}). In fact automated mitotic figure counting using deep learning in WSI has been explored, albeit with limited datasets and without practice-based evidence in clinical translation, limiting impact for broader conclusions on the value of the technology in clinical environments (\cite{aubreville2019field}, \cite{aubreville2020deep}, \cite{rao2018mitos}, \cite{balkenhol2019deep}). 

Clearly, there is great potential for leveraging deep learning technologies in digital pathology tasks, especially with the rising incidence in cancer globally and shortage of anatomic pathologists both in human and veterinary care (\cite{metter2019trends}, \cite{cockerell2013acvp}). However, despite several pre-market approvals, clinical deep learning applications have not been vigorously validated for reproducibility, generalizability and suffer from a lack of practice-based evidence (\cite{maddox2019questions}). This is most striking as automated histopathology analysis tools represent a significant opportunity for practice-based efficiencies given the time and effort consumed by identifying the area of highest mitotic figure density on a given slide as well as the task of counting the mitotic figures in that area. Both tasks are highly amenable to automation, warranting further exploration in real-world practice applications. Additionally, the subjective nature and inaccuracy of human performance of both tasks are well described (\cite{aubreville2020deep},\cite{bertram2020computerized},\cite{aubreville2019field},\cite{voss2015mitotic},\cite{veta2016mitosis}\cite{bonert2017mitotic}), highlighting the potential for improving accuracy with automation. 

The purpose of this study is to explore the development of this  deep learning based system to automate detection and quantification of mitotic figures on H\&E stained whole slide digital images not excluding any tumor types and describe the mechanism and subsequent impact of model deployment in a high throughput (thousands of WSI's per day from several thousand clinics over the course of a year) multi-site veterinary diagnostic pathology service. 

\section{Data}

Data collection was structured to support two distinct sub-tasks that the mitotic counting AI system needed to perform:
\begin{itemize}
    \item slide classification, to determine if mitotic count should occur
    \item mitotic figure counting
\end{itemize}

Data sampling for sub-tasks was performed independently. 

\subsection{Biopsy Slide Classification Data}
\label{sec:countNocountData}
For the biopsy slide classification, we ingested from a labeling tool in the workflow of pathologists. We collected 3845 slides of which $\approx 20\%$  were used as validation. The exact break down of slides is displayed in Table \ref{tab:biopsy_class_data}. The images are $1024\times$1024 slide thumbnails in their minimal magnification. For training, the images were resized to a size of $224\times$224.

\begin{table}[H]
\centering
\begin{tabular}{|l|l|l|l|l|}
\hline
\textbf{Category} & \textbf{Train} & \textbf{Validation} & \textbf{Image Example} \\  \hline 
(Count)                        & 1271  & 317 &  \raisebox{-.5\height}{\includegraphics[scale=0.25]{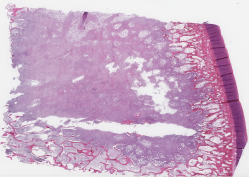}} \\ \hline 
(No-Count) Lipomas             & 703   & 176 & \raisebox{-.5\height}{\includegraphics[scale=0.25]{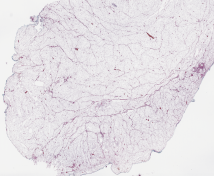}} \\ \hline 
(No-Count) Non-neoplastic lesions & 400   & 100  & \raisebox{-.5\height}{\includegraphics[scale=0.20]{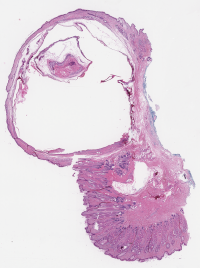}}\\ \hline 
(No-Count) Skin punch biopsies        & 180   & 45 & \raisebox{-.5\height}{\includegraphics[scale=0.1]{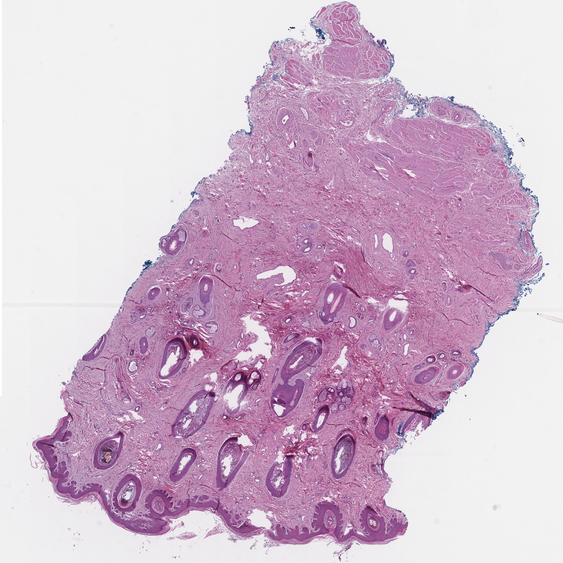}}   \\ \hline 
(No-Count) Skin\&subcutaneous tissues          & 522   & 131 & \raisebox{-.5\height}{\includegraphics[scale=0.25]{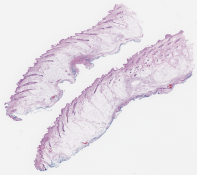}}  \\ \hline 
\end{tabular}
\caption{Biopsy slide classification data}
\label{tab:biopsy_class_data}
\end{table}

\subsection{Mitotic figure data collection}
 To build training sets of mitotic figures, board certified veterinary anatomic pathologists evaluated both snippets of varying sizes and WSI's of hematoxylin and eosin-stained slides, identifying mitotic figures by the cellular morphologic features of the various phases of mitosis as described in \cite{donovan2020mitotic}.  Mitotic figures were annotated for inclusion in training sets. Mitotic-like figures were excluded from the labeling process and thus were not included in training sets.  
 
 Training sets of mitotic figures are labeled often only for location (\cite{veta2016mitosis} ,\cite{bertram2019large}). Dependent on the modeling approach the labels are then represented as one our multiple pixel in the middle of the mitotic figure or bounding boxes. We instead used the full pixel masks of mitotic cells. This labor intensive labeling method was chosen because it allows for greater information flow in training and easy to implement generative approaches for data augmentation(\cite{isola2017image}). 
 
 In the first phase of the project we labeled 54 Mitotic Figures on 11 WSI's and extracted image patches of $150\times150$ Pixels with those mitotic figures in the middle. We also extracted 156 image patches with tissue without mitotic figures. We used this data to train a first model ( (\cite{ronneberger2015u}) to predict mitotic figures on 74 WSI's. We sub-sampled 1000 image patches with predicted mitotic figures in the middle as mitotic figures and labeled them. We used this data to train the next model and re-iterated the process.  To prevent a data distribution with small support we later added $600\times600$ and $1200\times1200$ pixel patches. 
 
 Targeted data sampling was applied by training an initial model on a labeled set of 54 Mitotic Figures on 11 WSI's. We also extracted 156 image patches with tissue without mitotic figures. We used this data to train a first model ( (\cite{ronneberger2015u}) to predict mitotic figures on a larger set of WSI's that we used as candidates for the next round of labeling. Each of the following iterations consisted of model critique and potential re-design, training, prediction of mitotic figure candidates and labeling of those candidates. 
 
 The initial phases of annotation used image patches of $150\times150$ pixels with mitotic figure candidates in the middle that pathologists would label making a binary decision. As the models improved the snippet sizes were increased to $600\times600$ and ultimately to $1200\times1200$ with potentially multiple mitotic figures inside each snippet in any location to prevent the exclusion of false positives from earlier model iteration in the data distribution. 
 The annotators were presented with a side by side copy of the image with the proposed mitotic figures masked on one copy (greyscale side to maximise contrast) of the image to show where the model believed they were. If the identified mitotic figures were all correct, a label of correct was applied, if all the predicted figures were wrong, a label of incorrect was applied and if their was a mix of correct and incorrect, the pathologists would mark on the unmasked image where the mitotic figures were. These marked images were then sent to have masks applied to them on the correct figures. The data was batched up and added to the training set. 
 \begin{figure}[H]
\centering
\includegraphics[scale=0.18]{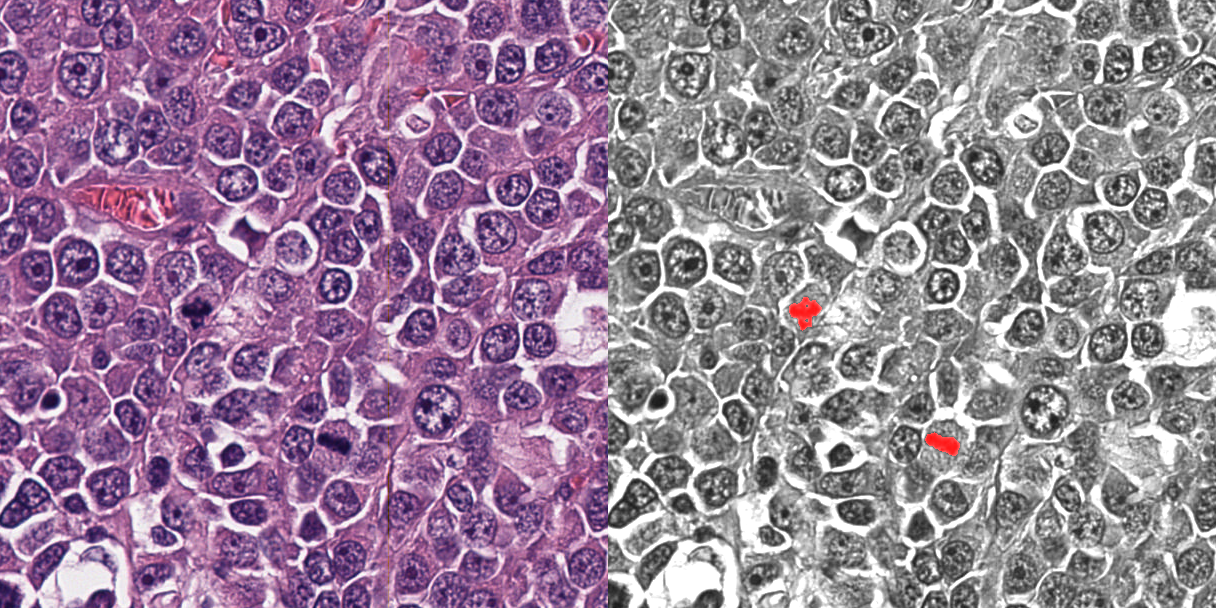}
\caption{Image Example of annotations presented to pathologists to create training data}
\label{fig:pathsMaskAnnotation}
\end{figure}
 
 In addition to the snippet evaluation, a set of 3 WSI's was fully labelled by three pathologists. The ground truth labels were defined as the union of all three pathologists' labels. The WSI's were lymphoma (canine), soft tissue sarcoma (feline), and mammary carcinoma (feline) cancer types and this formed the ground truth against which models were evaluated to assess performance.

\begin{figure}[H]
\includegraphics[scale=0.50]{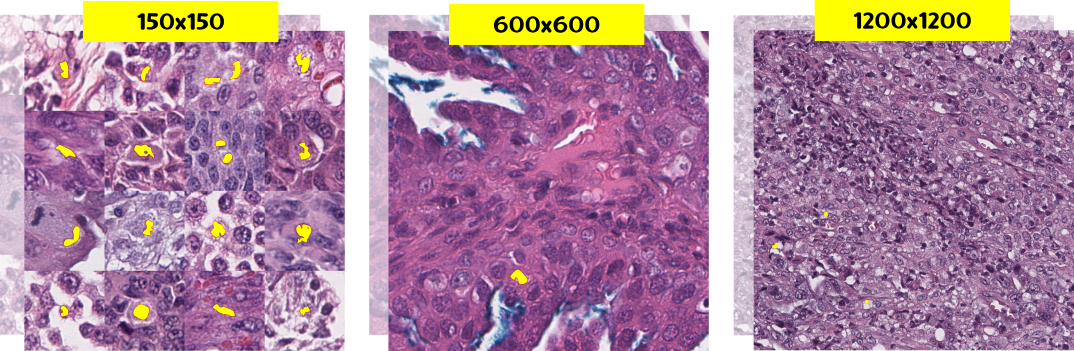}
\caption{Image Patches and Masks of $150\times150$, $600\times600$ and $1200\times1200$ pixels}
\label{fig:maskPatches}
\centering
\end{figure}
\section{Methods}

\subsection{System Overview}
\label{sec:inference}
One of the challenges of this project is to be able to deliver the result of the mitotic counting to all the pathologists in time to ensure they can utilize the information effectively whilst aiming for performance and optimizing the use of the computing resources. We now describe the system architecture that allows the use of the deep learning models described in this chapter to be utilized in an operational large scale multi site clinical practice. To do that we built and utilized  a pipeline composed of  multiple steps to pre- and post-process slides so as to be as resource efficient as possible. OncoPetNet is deployed in production and designed to operate on all scans that are uploaded in a real time fashion. We leverage a pull-based asynchronous architecture that is able to handle the high variation of scan uploads during a day.  When new scans are available they get added to a processing queue. When resources in the inference cluster are available, the slide gets downloaded and the following steps are performed: 

\begin{enumerate}
    \item The WSI is downloaded to the local machine. A $224\times224$ thumbnail is extracted. 
    \item The Biopsy Slide Classification is executed. If the Classifier returns 'no-count' the process is stopped and a no count status written to the database. If it returns 'count' it  is continued in the next step
    \item A tissue detection algorithm is executed. This non-model-based algorithm detects the foreground (tissue) and calculates coordinates for the mitotic figure detector.
    \item The Mitotic Figure Detector model is performed on all the coordinates. 
    \item After the pixel mask is predicted,  morphological filters are used to enhance predictive performance. A non-model-based algorithm is used to calculate coordinates of the Mitotic Figures from the enhanced pixel masks.   
    \item A \textit{k}-d tree based algorithm is used to find the 10HPF with the most Mitotic Figures.
    \item Coordinates for Mitotic Figures and the 10HPF are written into a XML and the results are written into a data base.

\end{enumerate}

\begin{figure}[H]
\centering
\includegraphics[scale=0.70]{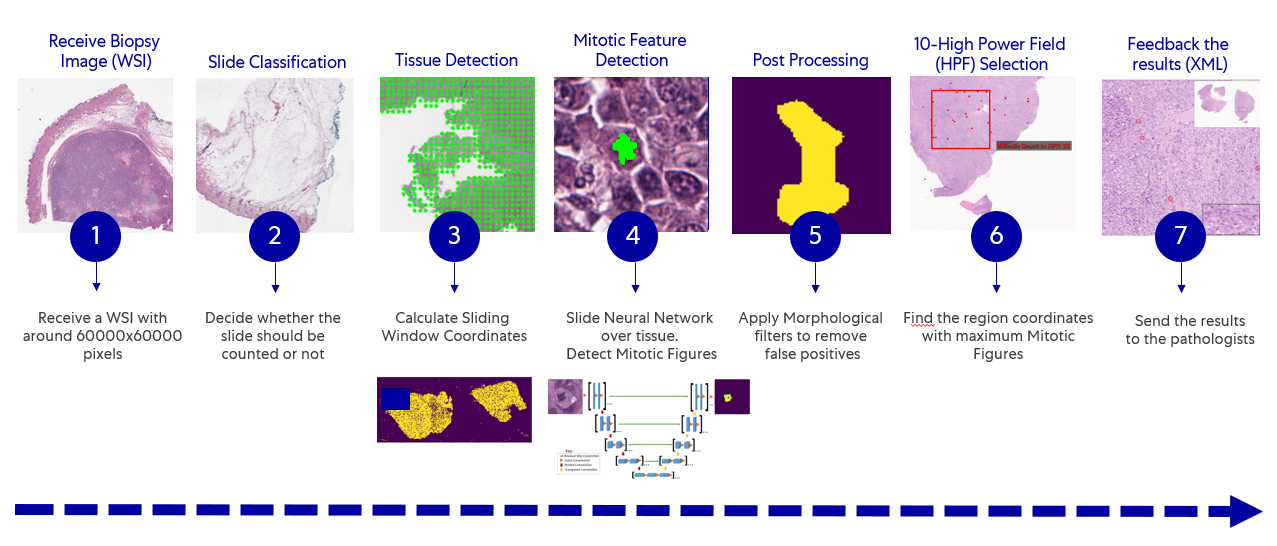}
\caption{Overview of the steps involved in our production system}
\label{fig:system}
\end{figure}

An overview of the system is displayed in Figure \ref{fig:system}. Deep Learning models were built in Pytorch 1.7 (\cite{paszke2019pytorch}).
We present all methods used in this chapter. 

\subsection{Biopsy Slide Classification}
\begin{figure}[H]
\centering
\includegraphics[scale=0.35]{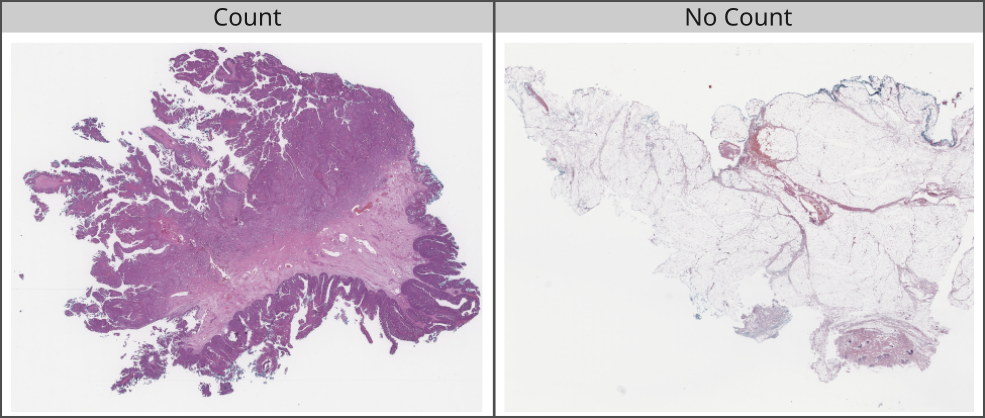}
\caption{Example of count and no count slides}
\label{fig:countNoCount}
\end{figure}

A biopsy slide classifier was created to pick only the relevant slides in a study. There are biopsy slides in which the amount of tissue is not relevant for the diagnosis, as you can see in the figure \ref{fig:countNoCount}. In these cases, the algorithm should skip the slide to save computing resources and focus on the slides that are important for the mitotic figure counting. A biopsy slide classification model was trained using a convolutional neural network approach. A pre-trained ImageNet \cite{deng2009imagenet} residual neural network (ResNet) \cite{DBLP:journals/corr/HeZRS15} architecture was used as backbone for the feature extraction followed by a fully connected layer classifier. Since inference time is an important factor for the project, a ResNet-18 (18 layers deep) was chosen based on trade-off between computing requirements and performance. The network has an image input size of $224\times$224 and was trained on 150 epoches with 3076 images distributed on Count and NoCount classes as demonstrated on \ref{sec:countNocountData}. The training was set using ADAM as the optimizer and CrossEntropyLoss for the loss function.

\subsection{Tissue Detection}
Coarse tissue segmentation supports the setting of back-lit digital pathology scanners, so the background is bright and lighten with white light source. Coarse tissue segmentation is performed on low resolution view of the whole slide image. Initially the low resolution image is resized to match the mitotic figure detector's sliding window size and high resolution image size, so that mitotic figure detector sliding window count is a whole number in high resolution image width. The low resolution image is then converted to LAB color space image and binary image threshold is applied on image L channel with OTSU threshold detection. Resulted binary image dilation, erosion and blurring performed to remove noise and refine the tissue boundaries.
\subsection{Mitotic Figure Detector}
\label{sec:unet}

Detection of mitotic figures with convolutional neural networks as the dominant algorithmic approach in an academic setting is well established (\cite{cirecsan2013mitosis}). These classification based methods allow for flexible training regimes and can lead to highly accurate models. For every $n$-th (with $n>0$ close to 1) Pixel on a WSI these models classify a patch of the WSI. This technique is very resource intensive in inference and therefore not feasible for real life lab situations.   The need for fewer sliding windows at inference time made end-to-end detection models arise naturally as potential solution. We trained different region based CNN's (\cite{ren2015faster})  but found that they could not reach the accuracy of our final system, which consists of an ensemble of three fully convolutional models at an equal inference time.  Instead of modeling the detection of Mitotic Figures directly we formulate the problem as a segmentation task where a deep learning model is used, combined with a deterministic post-processing methods. This approach allows for fast inference and very flexible model training. 
Our deep learning model consists of three fully convolutional neural networks following the encoder-decoder architecture with skip-connections (U-Net) described in in \cite{ronneberger2015u}. As Encoders we use Efficientnet-b5, Efficientnet-b3 (\cite{tan2019efficientnet}) and SE-Resnext(\cite{hu2018squeeze}, \cite{Xie2016}) that were pre-trained with ImageNet (\cite{deng2009imagenet}. We train our currently best model on $1790$ patches of $600 \times 600$ pixels ( $\mathscr{D}_{600}$) and $6359$ patches of $150 \times 150$ pixels ( $\mathscr{D}_{150}$). We aim to minimize 

\begin{align}
    \alpha \cdot \mathscr{L}(\theta, \mathscr{D}_{150}) + (1-\alpha) \cdot \mathscr{L}(\theta, \mathscr{D}_{600})
\end{align}
Where $ \mathscr{L}$ is a loss function and $\alpha \in [0,1]$ is a hyper parameter. Our currently best model was trained on 50 epochs where $\mathscr{L}$ was set as BCE-soft-Dice-Loss (\cite{isensee2018nnu}) and subsequently 100 epochs for which we switched to soft Dice Loss (\cite{sudre2017generalised}). We validate our model against $448$ images of $600 \times 600$ pixels and chose the model parameters after the epoch that has the highest f1-score on the validation set. 

We combine the three resulting models by averaging their predictions at runtime.

\subsection{Pixel Mask Post-Processing}
Following geometrical and morphological procedures are performed to refine inference results, reduce false positives and improve precision of 10HPF region location.

1. Preparation, mitotic figure instance segmentation.

Initial binary mask is a result of MF detector semantic segmentation and it's size equal to the model's output tensor size e.g. $600\times600$ pixels.  The mask may contain zero or more mitotic figures. We are further concerned to gain metric information on each instance of mitotic figures. MF instance segmentation is performed by computing connected components of given binary mask which labels connected pixels with unique integer label. So pixels belong to candidate mitotic figure instance receive unique integer number.

2. Excluding small-sized Mitotic-like Figures

It was empirically evident that sometime MF detector model generates the mask of mitotic figure that does not meet biological standards. To avoid small mitotic-like figures from being counted as MFs, we empirically defined the size of 3 microns as a minimum acceptable size along mitotic figure width.

Task of excluding abnormally small-sized MFs is reformulated to following task: validate mitotic figures against given minimal acceptable width criteria on a given MF instance segmentation mask where MF could be rotated at arbitrary angle.

The algorithm finds minimum area bounding box around MF mask, assuming it's longest side to be the width of MF i.e. for every instance of mitotic figure find a rectangle of a minimum area enclosing the mask's approximated contour. The width is then measured as max(height, width) of rectangle.

Introducing min allowed MF width led to significant reduction in false positives and added +20

3. Counting mitotic figures at late stages of mitosis

Mitotic figures in anaphase and telophase appear as two separated dense linear aggregates at variable distances apart. In late telophase these aggregates may be separated by a thin strand of cytoplasm in preparation for cytokinesis, the process of separation into two identical daughter cells. Binary mitotic figure detector and following mitotic counting may double count these phases of mitosis.

To avoid double counting of anaphase and telophase MFs proposed solution is to combine two MFs distant within given interpolar (between daughter cells) max distance. Pathology group empirically defined 15 micrometers as the max interpolar distance.

The proposed solution: 1.Dilate binary mask with the number of dilation iterations as such that dilated mask gain in size half the max interpolar distance so all MF instances masks within the max interpolar distances would be combined into one instance mask. 2.Find connected components in dilated masks. 3.Label MF instances according to connected components markers.

The limitation of this solution is inherited of sliding algorithm local field of view the sliding window, so late stage MFs for which two daughter cells masks are separating into 2 different sliding windows may still be double counted. It is proposed to resolve this issue by global instance segmentation using \textit{k}-d tree and max interpolar distance neighbors search.

\subsection{\textit{k}-d tree based 10-HPF search}
In digital pathology 10HPF stands for 10 high power fields and is a standard square region for reporting mitotic count. We follow \cite{meuten2016mitotic} and use $\mathscr{}{2.37mm^2}$ as a metric size of 10HPF square region, which we then convert to pixel size using whole slide image SVS format property microns per pixel written by WSI scanners. When reporting mitotic count pathologists find a 10HPF with highest count of mitotic figures, hence our algorithm finds 10HPF area with the maximum MF count.

The problem was formulated as follows: given the set of detected Mitotic Figures represented by their minimum areas bounding rectangles' centers find a highest MFs count square 10HPF region. The problem can be re-formulated as a range search in $\mathbb{R}^2$. \newline Let $\mathbb{S}$ be a set of $N$ points in $\mathbb{R}^2$, for $\mathscr{MF}_{i}, {i}\in{[1,N]}$, $N$ is the total number of detected mitotic figures, find and count all MFs inside the 10HPF range with the centre at $\mathscr{C}_{j}$, ${j}\in{[1,N^2]}$, ${\mathscr{C}}\in{\mathbb{S_X}\times \mathbb{S_Y}+r}$, where $r$ is 10HPF radius $r=\frac{\sqrt{2.37}}{2}$mm

\textit{k}-d tree is an efficient binary tree data structure for k-dimensional search space proposed by Bentley \cite{bentley1975multidimensional}, \cite{compgeom} in the case of 2-dimensional search space it builds the tree data structure in $O(n\log{} n)$ which support handling range reporting queries in $O(\sqrt{n} +k)$, n - total number of points, k - number of reported points.
Our proposed solution for highest count 10HPF region with exact algorithm as follows:

\begin{enumerate}
\item Build \textit{k}-d tree for set of N 2D points (the centres of all detected Mitotic Figures), with $\mathscr{L}^{\infty}$ distance metric also known as Chebyshev's distance, so the distance between 2 points P1 and P2 in 2-dimensional space is defined as follows $D_{Chebyshev}(P1,P2)=max(|X_{P1}-X_{P2}|, |Y_{P1}-Y_{P2}|)$ A region in a space with Euclidean metric enclosed in a circle with radius r becomes region enclosed in a square with side of 2*r under Chebyshev's distance.
\item The set of all possible 10HPF regions' corners is a set derived from combination between X and Y coordinates of MF centers or as a cartesian product between MF centers' X and Y coordinates. In order to query points for $\mathscr{L}^{\infty}$ \textit{k}-d tree search query all possible 10HPF region corners are converted to their centers by adding 10HPF radius. For every 10HPF center proposal then query the \textit{k}-d tree built at step 1 for all neighbors in radius of 10HPF radius.
\item Report the region with the maximum count of neighbors.
\end{enumerate}
Our 10HPF search solution significantly improves the whole slide image processing time and for higher Mitotic Figure count slides it saves up to 90 seconds per slide on synthetically generated whole slide mitosis map, see Table \ref{tab:tenHPF_time_comparison} in Supplemental Data. During 30 days of practical system usage, the time improvement was 60 seconds per whole slide image in average across low and high mitotic count cases as reported by usage and performance logs.

\subsection{Hard Example Mining}

The model results are integrated in the pathologist workflow. The pathologist will receive the information from the model as annotations that are rendered from the generated XML file in a WSI viewer.

\begin{figure}[h]
\caption{A fully annotated whole slide image on the pathologist's viewer}
\label{fig:wsiannotated}
\includegraphics[scale=0.50]{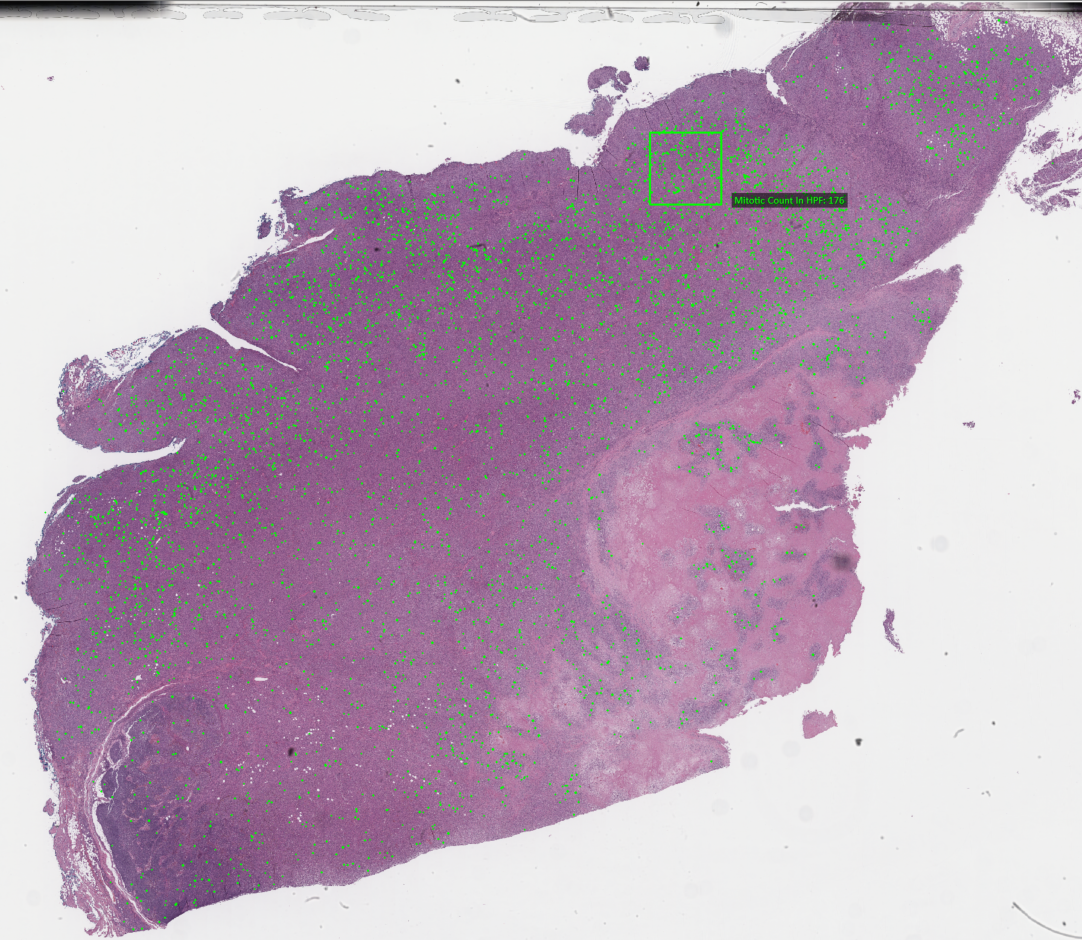}
\centering
\end{figure}

\begin{figure}[h]
\caption{Zoomed-in view of an annotated WSI}
\label{fig:wsiannotated}
\includegraphics[scale=0.50]{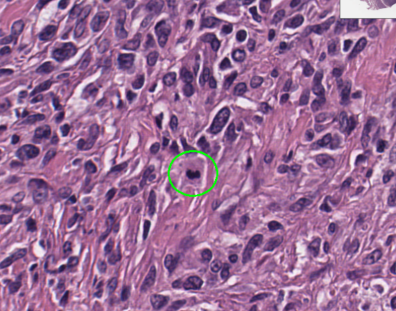}
\centering
\end{figure}

The pathologists can themselves annotate False Positive and False Negative Examples by adding annotations to their slide viewer. Those feedback annotations get stored in a database that is read out daily. Based on the feedback information new images for fine tuning of the Detection model (\ref{sec:unet}) are created. Fine-tuning is performed on a regular basis, dependent on the feedback volume. 

\section{Experiments}
\subsection{Metrics}

To determine the accuracy of the AI-determined mitotic count, a data set of 41 canine and feline cancer cases submitted to Antech Diagnostics for histopathologic evaluation were used. These cases represented 14 distinct cancer types with mitotic counts ranging from very low to very high.  In total, across the 41 cases there were 110 whole slide images.  Data from the original manual mitotic counts (non-AI assisted MC) on each slide were available and were performed under standard diagnostic pathology workflow conditions by two pathologists. The model then performed mitotic figure annotations and counts on all slides to obtain the AI-only MC.  Two board-certified veterinary pathologists were then tasked with independently evaluating the AI-annotated mitotic figures in the AI-preselected 10HPF containing the highest mitotic density, identifying false positives and false negatives, and determining an AI-assisted MC.  The non-AI assisted MC for each slide was then compared to the AI-only MC and the AI-assisted MC.    
\subsection{Inference}
We used internal Mars Inc. infrastructure to
measure the inference runtimes of the mitotic figure counting AI system. Inference time represents the time spent of steps 1-6 of the AI system (see Figure \ref{fig:system}). The experiments were performed on
hardware settings closely matching the production setup. Lower inference times are desirable.

\section{Workload and Deployment}
One of the big challenges on this work was creating a dynamic and fast system that could be capable of delivering the AI results to all the pathologists before they start working on their study cases.

As understanding the workload and time constraints were fundamental, we collected data from five months of study cases and measured the average number of slides received by the pathologists every day, as you can see in the table \ref{tab:worloadTable}.
The system should process around 3,323 slides a day, in which 1,163 requires the Mitotic Counting. Given that every slide has an average of 8,164 images (sliding window approach), it is about .5 million $600\times$600 images in the inference pipeline per day.

\begin{table}[H]
\centering
\begin{tabular}{|l|l|}
\hline
\textbf{} & \textbf{Production Slide Workload} \\ \hline
Total \# of slides (Count + NoCount)  & 48,354  \\ \hline
Avg \# of (Count + NoCount) slides & 3,323 slides/day  \\ \hline
Avg \# of (Count) slides  & 1,163 slides/day \\ \hline
Ratio (Count/No count) & 35\%  \\ \hline
Avg \# of images ($600\times$600) & 8,164/slide  \\ \hline
Avg \# of images ($600\times$600) per day & 9,44,732/(slide*day) \\ \hline

\end{tabular}
\caption{In Production Slide Workload}
\label{tab:worloadTable}
\end{table}

Delivering such a complex set of modules with the presented time constraint requires a robust system architecture and a series of optimization techniques were used for speeding-up the Mitotic Counting process.

\subsection{System Modules}

\begin{figure}[h]
\includegraphics[scale=0.6]{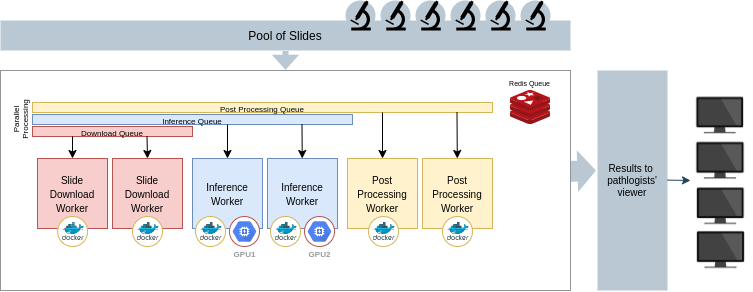}
\caption{Deployed Distributed architecture}
\label{fig:mfnetDeploy}
\centering
\end{figure}

The systems are distributed across the east and west coasts of the United States for reducing network latency on downloading the slides from the central repositories. Each coast contains five servers with a multi-GPU configuration. Aiming to extract the most of the GPU's and CPU's, a decentralized architecture was created using Redis Queues for caching the requests and docker containers to orchestrate and run the different modules, as you can see in the fig.\ref{fig:mfnetDeploy}. 

The asynchronous structure consists of three different scalable blocks: Download Worker, Inference Worker and Post-Processing Worker.

The download worker is responsible for downloading the slides from the network file system (NFS) and creating a buffer of slides in the local file-system, so the inference module can have access to the slides locally. The inference worker module receives the slides and process them to extract the Mitotic figures as described on steps from 1-4 on the Inference Engine section \ref{sec:inference}. The workload is balanced and shared across different workers based on each GPU availability.

Finally, the post processing module will run the final steps from 5-7 of the Inference Engine, providing the final results to the pathologist's viewer. 

The download and post processing workers are Network/Disk and CPU-bounded, so the processing bottlenecks are related to the network bandwidth and the number of CPU cores. On the other hand, the inference worker is GPU-bounded, meaning that the processing limit is related to the GPU cores and VRAM memory. Given that each module has different processing timing requirements to finish the task, separating their execution in asynchronous fashion enabled the system to run all of the steps at the same time on the same server, reducing the GPU idle time and optimizing the hardware resources.

Each class of worker has it's own Redis queue. Thus, scaling the solution is as simple as adding a new docker container with the respective worker module.

\subsection{Optimization techniques}
Different techniques were used to either increase the system throughput and decrease the inference time. Where possible we leveraged the optimiyation capabilities of Pytorch 1.7 (\cite{paszke2019pytorch})

Increasing the number of images loaded into the GPU (batch size) helps the GPU to execute less instructions per inference, improving the performance. One of the steps of the inference phase is the data preparation. The image has to be extracted from the slide (sliding window approach), normalized and reshaped to match the GPU input format. Thus, increasing the batch size also increases the data preparation time. Therefore, a multi-processing approach was used to load and pre-process multiple images in parallel, that combined with the increase of the batch size improved the overall inference performance by 4x. The inference module stacks 16 slide regions ($600\times$600 images) and transforms them into a vector with shape in the format [batchSize, numberOfChannels, imageWidth, imageHeight].

It is possible to spawn multiple workers in the same server, so the system can have more than one inference module per GPU, increasing the throughput. The number of workers depends on the amount of resources available on the system.

Moreover, reducing the data format to use half precision computation also boosts the inference performance. Half precision allows Float16 calculations instead of a full Float32 data type without significant accuracy loss.

Finally, the architecture has a multi-node support, so it's easy to scale the workers across different servers.

\section{Results}

\subsection{Test Metrics on fully annotated slides}
Fully annotated slides were used as test set. To prevent overfitting on the test set, we only ran inference on test set if a model set a new benchmark versus earlier models on the validation set. Thresholding was performed on validation data only, not on the test set. Results are displayed in table \ref{tab:3slidetest-table}.

\begin{table}[H]
\centering
\begin{tabular}{|l|l|l|l|}
\hline
                 & \textbf{Slide 1}   & \textbf{Slide 2}  & \textbf{Slide 3}   \\ \hline
Precision OncoPetNet & 0.725570  & 0.596244 & 0.86658   \\ \hline
Recall OncoPetNet   & 0.8583815 & 0.581032 & 0.829066  \\ \hline
F1 OncoPetNet       & 0.7864078 & 0.60012  & 0.8474119 \\ \hline
\end{tabular}
\caption{Results on three test slides. Slide 1 is a canine lymphoma. Slide 2 is a feline soft tissue sarcoma. Slide 3 is a feline mammary carcinoma.  }
\label{tab:3slidetest-table}
\end{table}    

\subsection{Deployment Result}

The combination of a highly scalable architecture and the applied optimization methods resulted on a 0.27 min/slide rate on a multi-processing and multi-node environment with ten GPU-powered servers. It is 24x faster when comparing to a single node and single inference worker inference time. Thus, the asynchronous approach not only optimize the usage of hardware resources, but also gives the flexibility to increase the computing power in a fast-pace.

\begin{table}[H]
\centering
\begin{tabular}{|l|l|}
\hline
\textbf{Metric} & \textbf{Value} \\ \hline
Avg compute time (single node) & 6.5min/slide \\ \hline
Avg compute time (multi-node) & 0.27min/slide \\ \hline

\end{tabular}
\caption{Performance metrics in an optimized multi-process and multi-node environment.}

- our solution is better because ...
- integrating with workflow
- live full production lab not academic group
- scalability part 

\label{tab:worload-and-performance}
\end{table}

\subsection{Accuracy of the AI-determined count}
\label{sec:acc_prod}
Histopathologic review of the 41 case data set identified significant differences between the Non-AI assisted MC, AI-only MC, and the AI-Assisted MC.  In general, across all tumor types the AI-only MC was higher than the Non-AI assisted MC.  Review of the automated results by pathologists generally resulted in an AI-assisted MC slightly lower than the AI-only MC, but in general the AI-assisted MC remained higher than the Non-AI assisted MC across all tumor types.  Of the 41 cases, 32 were cancer types with grading schemes, 7 of which experienced an increase in grade with both the AI-only and AI-assisted MC as compared to the Non-AI assisted MC (see Table \ref{tab:grade_increase}). The results for the entire data set are found in Figure \ref{fig:Mitotic_index_determination_with_and_without_AI_assistance} in Supplemental Data.  

\begin{table}[H]
\centering
\begin{tabular}{|l|l|l|l|l|l|}
\hline
\textbf{Case \#} & \textbf{Tumor Type} & \textbf{\begin{tabular}[c]{@{}l@{}}Non-AI \\ assisted MC\end{tabular}} & \textbf{\begin{tabular}[c]{@{}l@{}}AI-only \\ MC\end{tabular}} & \textbf{\begin{tabular}[c]{@{}l@{}}AI Assisted \\ MC\end{tabular}} & \textbf{\begin{tabular}[c]{@{}l@{}}Change in \\ Grade\end{tabular}} \\ \hline
3                & Mammary carcinoma, feline   & 30                                                                     & 129                                                            & 117                                                                & 2 → 3                                                               \\ \hline
13               & Mammary carcinoma, canine   & 11                                                                     & 34                                                             & 25                                                                 & 1 → 2                                                               \\ \hline
28               & Mammary carcinoma, canine   & 8                                                                      & 17                                                             & 13                                                                 & 2 → 3                                                               \\ \hline
32               & Mast cell tumor, canine     & 4                                                                      & 26                                                             & 13                                                                 & Low → High                                                          \\ \hline
7                & Soft tissue sarcoma, canine & 3                                                                      & 15                                                             & 11                                                                 & 1 → 2                                                               \\ \hline
39               & Soft tissue sarcoma, canine & 3                                                                      & 90                                                             & 69                                                                 & 2 → 3                                                               \\ \hline
40               & Soft tissue sarcoma, canine & 3                                                                      & 63                                                             & 43                                                                 & 2 → 3                                                               \\ \hline
\end{tabular}
\caption{Selected results on comparing mitotic counts determined by pathologists without AI assistance (Non-AI assisted MC), by AI alone (AI-only MC), and by pathologists with AI assistance (AI-Assisted MC) with the resulting changes in grading.}
\label{tab:grade_increase}
\end{table}

\subsection{Statistical Analysis}

We conducted a statistical analysis of the influence of the different methods in section \ref{sec:acc_prod} on the mitotic count. For each method $i$ (AI-Assisted, AI-Only and Non Assisted) and each cancer type $j$ (see Figure \ref{fig:Mitotic_index_determination_with_and_without_AI_assistance}) we model the mitotic count $y_{ij}$ using a Poisson regression 

\begin{align}
\label{eq:regression}
y_{ij} &\sim Poisson(e^{\lambda_i + \beta_j}) \\
\beta_j &\sim  student\_t(3, 0, 2.5) 
\end{align}
where the $\lambda_i$'s control for method used to find the mitotic count and $\beta_j$'s adjust for variation among cancer types. We present the coefficients $\exp(\lambda_i)$ in table \ref{tab:stat_results} which represent the relative mitotic count after controlling for cancer type

\begin{table}[H]
\centering
\begin{tabular}{|l|l|l|l|}
\hline
\textbf{AI Assistance} & \textbf{Relative mitotic count}        \\ \hline
Non Assisted  & 12.192 (9.650, 15.522) \\ \hline
Assisted      & 35.835 (28.508, 45.404) \\ \hline
AI Only       & 43.935 (34.917, 55.446) \\ \hline
\end{tabular}
\caption{Estimated Mean for three methods of counting $\exp(\lambda_i)$. 0.95 Credibility Intervals in parenthesis}
\label{tab:stat_results}
\end{table}

\begin{figure}[h]
\caption{Conditional $\lambda_i$ of Poisson regression (\ref{eq:regression})}
\label{fig:stats}
\includegraphics[scale=0.2]{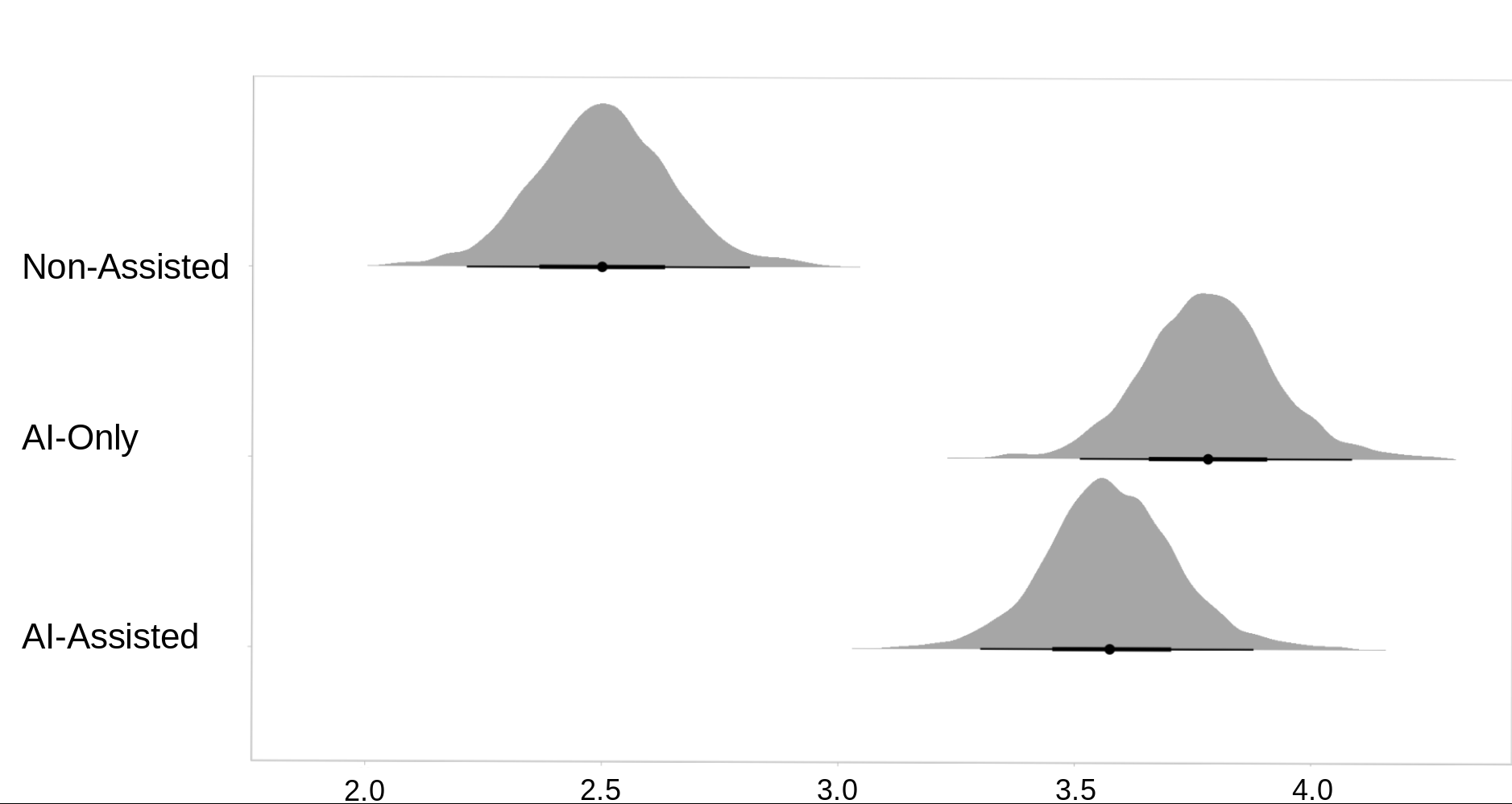}
\centering
\end{figure}

We display the full posteriors for the three main effects in \ref{fig:stats} on log-scale for readability purposes.

\section{Discussion}

The purpose of this study was to explore the development of an automated deep learning based system to automate detection and quantification of mitotic figures on H\&E- stained whole slide digital images and describe the mechanism and subsequent impact of model deployment in a high throughput multi-site veterinary diagnostic pathology service. We found that expert-level deep learning based models for automating important diagnostic tasks can be successfully achieved, and further, deployed on a daily average of 3,323 Whole Slide Images within the workflow and time constraints of a high throughput multi-site diagnostic pathology service. This work has important implications for the development and deployment of automated deep-learning systems in clinical histopathology practices, and informs best practices for translating deep learning models into digital pathology practices.

\paragraph{}

 Application of high-performance deep learning systems at scale may provide critical insights that can serve to address the gap in translation for these promising technologies for both veterinary and human pathology diagnostics into clinical practice. In diagnostic pathology, mitotic count determination can be critical in determining whether a neoplasm is benign or malignant, as well as providing valuable prognostic information. To date deployment of AI systems have been limited despite that two key steps of the traditional manual mitotic count could be greatly improved by automation with high-performance deep learning models in digital pathology: site selection and mitotic figure quantification. Prior work to increase efficiency and accuracy of mitotic figure detection and quantification has been attempted a number of ways.  Marking mitotic figures with immunohistochemical (IHC) stains, such as Phosphohistone H3, has been shown to reduce time of counting as well as improve inter-observer variability (\cite{voss2015mitotic}. Going even further, another model labeled mitotic figures with Phosphohistone H3 and developed an automated computational detection system that identified and quantified the labeled mitotic figures(\cite{puri2019automated}.  However, in both models IHC staining adds additional time and cost to histopathologic evaluation and is impractical as a standard cancer diagnostic approach in veterinary medicine. In other models, site selection and mitotic figure quantification were automated and resulted in improved efficiency and accuracy as compared to pathologists, however, these models required manual annotation of mitotic figures prior to site selection and quantification (\cite{bertram2020computerized}\cite{aubreville2020deep}\cite{pantanowitz2020accuracy}). Automated models to detect mitotic figures in H\&E slides have been developed by training on static datasets (\cite{challenge2016tupac16}) and even comparatively large data sets are drawn from a relative small number of known cases of single tumor types (\cite{bertram2019large}). In a real world lab production setting, models must be able to perform against unknown cases in a variety of laboratories and encompass challenges around staining variation, slide scanning softwares, and across numerous tumor types; each of these real-world barriers leads to covariate shift in the clinical deployment environment demanding the collection of training data in a real life lab setting framed as a dynamic process that demands consideration of real-world data best suited to represent the data distribution mentioned.  We found that the collection process also needs to be adaptable to the prediction profile of the current model iteration as well as resource conscious.

\paragraph{}
Most of the works in literature focus on creating tools for helping the research community to easily handle and interact with WSI's\cite{8698753},\cite{pedersen2020fastpathology}. However, driving a high scale automated system for production requires dealing with complexities that are not present on a scientific environment. This study focuses on deploying the mitotic counting ecosystem along with an existent large diagnostic lab setting using a robust architecture that delivers speed, scalability and resilience. The system account for a variation of workload and time constraints imposed by a daily work routine of the pathologists. The asynchronous architecture gives independence for each one of the nodes, making possible to scale the computing power with minor or no software changes. Also, the distributed workers are failure proof, if one node stop working, the others remain live, which also reduces the system downtime on maintenance windows.

\paragraph{}

To date there is little literature regarding the impact of automated processes and impact in downstream clinical decision-making. An analogous circumstance occurred during the rapid adoption of automated blood cell counting, prior to which a human eye-count differential was the only way to identify cell types and their relative proportion for nearly 100 years.  The rapid adoption of automated systems due to improved accuracy, consistency, and efficiency led to immediate impact on downstream clinical decision making (\cite{pierre2002peripheral}); in fact the traditional review of all automated hematology instrument results by preparation, staining, and microscopic examination of a blood film has disappeared in modern healthcare. Adjustments had to be made in clinical practice as a result, given historical considerations of patient grade and prognosis had been based on imperfect human approaches, drawing important parallels to this study. While this work was not designed explicitly to evaluate impact on downstream changes to clinical staging, we found that nearly 22\% of the cancer cases in our data set that have a defined grading scheme, with use of the automated systems would have been clinically up-staged, with significant implications on clinical decision making and therapeutic decisions. This is perhaps the most significant impact to clinical care by automated systems (with or without human expert oversight) as new grading considerations may need to be considered when an AI system is used in the grading, particularly for certain cancer types, given that upstaging cancers could fundamentally change prognosis based on existing clinical data and even adversely alter patient outcomes (i.e. euthanasia/comfort care vs aggressive therapy). Future work focused on clinical impact our outcomes will need to take the use of automated systems into account prior to treatment recommendations.

\paragraph{}
This study has several important limitations.  Mitotic figures identified by the model were not restricted to cancer tissue alone, resulting in normal parenchymal tissue mitotic activity factoring into the AI-only MC.  This is an important consideration in tumors with a low mitotic count, and requires pathologist review to account for non-tumor mitotic figure counting, underscoring the importance of having humans involved in interpretation of AI-generated MCs. Further work to restrict mitotic figure counting to tumor-only tissue will improve accuracy of the AI-only MC.  Clinical scoring was performed retrospectively and further prospective work with OncoPetNet in practice may find mitotic figure count and time variance as part of routine practice. The clinical impact of the change in stage for several cancer types via the automated counting was out of scope for this work, but as discussed likely has important downstream implications on clinical decision making. Further work could include prospective studies that allow clinicians to evaluate cases with and without the algorithm in real time, allowing them to collect and consider this information in their ultimate diagnosis and allow for a more accurate comparison to true clinical practice.  Performance metrics for the model deployment workflow were not evaluated as a direct comparison to the pre-deployment practice; future work can explore the economic and clinical efficiencies resulting from the model deployment and examine sustained performance over time.  

\paragraph{}
In conclusion, this work represents the first successful automated deployment of deep learning systems for real-time expert-level performance on important histopathology tasks at scale in a high volume clinical practice. The resulting impact outlines important considerations for model development, deployment, impact on clinical decision making, and informs best practices for implementation of deep learning systems in digital histopathology workflows.

\section{ACKNOWLEDGMENTS}
The authors would like to thank Francisco Massa who helped us with valuable advice to leverage Pytorch's capabilities.

\bibliographystyle{plain}
\bibliography{references}

\begin{thebibliography}{10}

\bibitem{aubreville2019field}
Marc Aubreville, Christof~A Bertram, Christian Marzahl, Corinne Gurtner,
  Martina Dettwiler, Anja Schmidt, Florian Bartenschlager, Sophie Merz, Marco
  Fragoso, Olivia Kershaw, et~al.
\newblock Field of interest prediction for computer-aided mitotic count.
\newblock {\em arXiv. org}, (1902.05414), 2019.

\bibitem{aubreville2020deep}
Marc Aubreville, Christof~A Bertram, Christian Marzahl, Corinne Gurtner,
  Martina Dettwiler, Anja Schmidt, Florian Bartenschlager, Sophie Merz, Marco
  Fragoso, Olivia Kershaw, et~al.
\newblock Deep learning algorithms out-perform veterinary pathologists in
  detecting the mitotically most active tumor region.
\newblock {\em Scientific RepoRtS}, 10(1):1--11, 2020.

\bibitem{balkenhol2019deep}
Maschenka~CA Balkenhol, David Tellez, Willem Vreuls, Pieter~C Clahsen, Hans
  Pinckaers, Francesco Ciompi, Peter Bult, and Jeroen~AWM van~der Laak.
\newblock Deep learning assisted mitotic counting for breast cancer.
\newblock {\em Laboratory investigation}, 99(11):1596--1606, 2019.

\bibitem{bejnordi2017diagnostic}
Babak~Ehteshami Bejnordi, Mitko Veta, Paul~Johannes Van~Diest, Bram
  Van~Ginneken, Nico Karssemeijer, Geert Litjens, Jeroen~AWM Van Der~Laak,
  Meyke Hermsen, Quirine~F Manson, Maschenka Balkenhol, et~al.
\newblock Diagnostic assessment of deep learning algorithms for detection of
  lymph node metastases in women with breast cancer.
\newblock {\em Jama}, 318(22):2199--2210, 2017.

\bibitem{bentley1975multidimensional}
Jon~Louis Bentley.
\newblock Multidimensional binary search trees used for associative searching.
\newblock {\em Communications of the ACM}, 18(9):509--517, 1975.

\bibitem{compgeom}
Mark~de Berg, Otfried Cheong, Marc~van Kreveld, and Mark Overmars.
\newblock {\em Computational Geometry: Algorithms and Applications}.
\newblock Springer-Verlag TELOS, Santa Clara, CA, USA, 3rd ed. edition, 2008.

\bibitem{bertram2020computerized}
Christof~A Bertram, Marc Aubreville, Corinne Gurtner, Alexander Bartel, Sarah~M
  Corner, Martina Dettwiler, Olivia Kershaw, Erica~L Noland, Anja Schmidt,
  Dodd~G Sledge, et~al.
\newblock Computerized calculation of mitotic count distribution in canine
  cutaneous mast cell tumor sections: mitotic count is area dependent.
\newblock {\em Veterinary pathology}, 57(2):214--226, 2020.

\bibitem{bertram2019large}
Christof~A Bertram, Marc Aubreville, Christian Marzahl, Andreas Maier, and
  Robert Klopfleisch.
\newblock A large-scale dataset for mitotic figure assessment on whole slide
  images of canine cutaneous mast cell tumor.
\newblock {\em Scientific data}, 6(1):1--9, 2019.

\bibitem{bonert2017mitotic}
Michael Bonert and Angela~J Tate.
\newblock Mitotic counts in breast cancer should be standardized with a uniform
  sample area.
\newblock {\em Biomedical engineering online}, 16(1):1--8, 2017.

\bibitem{challenge2016tupac16}
Tumor Proliferation~Assessment Challenge.
\newblock Tupac16-miccai grand challenge, 2016.

\bibitem{cirecsan2013mitosis}
Dan~C Cire{\c{s}}an, Alessandro Giusti, Luca~M Gambardella, and J{\"u}rgen
  Schmidhuber.
\newblock Mitosis detection in breast cancer histology images with deep neural
  networks.
\newblock In {\em International conference on medical image computing and
  computer-assisted intervention}, pages 411--418. Springer, 2013.

\bibitem{cockerell2013acvp}
Gary Cockerell, Curtis Colleton, Claire Andreasen, and Thomas Monticello.
\newblock The acvp/stp coalition for veterinary pathology fellows adapts to
  changing employment demographics.
\newblock {\em Toxicologic pathology}, 41(1):137--138, 2013.

\bibitem{deng2009imagenet}
Jia Deng, Wei Dong, Richard Socher, Li-Jia Li, Kai Li, and Li~Fei-Fei.
\newblock Imagenet: A large-scale hierarchical image database.
\newblock In {\em 2009 IEEE conference on computer vision and pattern
  recognition}, pages 248--255. Ieee, 2009.

\bibitem{dennis2011prognostic}
MM~Dennis, KD~McSporran, NJ~Bacon, FY~Schulman, RA~Foster, and BE~Powers.
\newblock Prognostic factors for cutaneous and subcutaneous soft tissue
  sarcomas in dogs.
\newblock {\em Veterinary Pathology}, 48(1):73--84, 2011.

\bibitem{donovan2020mitotic}
Taryn~A Donovan, Frances~M Moore, Christof~A Bertram, Richard Luong, Pompei
  Bolfa, Robert Klopfleisch, Harold Tvedten, Elisa~N Salas, Derick~B Whitley,
  Marc Aubreville, et~al.
\newblock Mitotic figures—normal, atypical, and imposters: A guide to
  identification.
\newblock {\em Veterinary Pathology}, page 0300985820980049, 2020.

\bibitem{esteva2017dermatologist}
Andre Esteva, Brett Kuprel, Roberto~A Novoa, Justin Ko, Susan~M Swetter,
  Helen~M Blau, and Sebastian Thrun.
\newblock Dermatologist-level classification of skin cancer with deep neural
  networks.
\newblock {\em nature}, 542(7639):115--118, 2017.

\bibitem{gulshan2016development}
Varun Gulshan, Lily Peng, Marc Coram, Martin~C Stumpe, Derek Wu, Arunachalam
  Narayanaswamy, Subhashini Venugopalan, Kasumi Widner, Tom Madams, Jorge
  Cuadros, et~al.
\newblock Development and validation of a deep learning algorithm for detection
  of diabetic retinopathy in retinal fundus photographs.
\newblock {\em Jama}, 316(22):2402--2410, 2016.

\bibitem{DBLP:journals/corr/HeZRS15}
Kaiming He, Xiangyu Zhang, Shaoqing Ren, and Jian Sun.
\newblock Deep residual learning for image recognition.
\newblock {\em CoRR}, abs/1512.03385, 2015.

\bibitem{hu2018squeeze}
Jie Hu, Li~Shen, and Gang Sun.
\newblock Squeeze-and-excitation networks.
\newblock In {\em Proceedings of the IEEE conference on computer vision and
  pattern recognition}, pages 7132--7141, 2018.

\bibitem{isensee2018nnu}
Fabian Isensee, Jens Petersen, Andre Klein, David Zimmerer, Paul~F Jaeger,
  Simon Kohl, Jakob Wasserthal, Gregor Koehler, Tobias Norajitra, Sebastian
  Wirkert, et~al.
\newblock nnu-net: Self-adapting framework for u-net-based medical image
  segmentation.
\newblock {\em arXiv preprint arXiv:1809.10486}, 2018.

\bibitem{isola2017image}
Phillip Isola, Jun-Yan Zhu, Tinghui Zhou, and Alexei~A Efros.
\newblock Image-to-image translation with conditional adversarial networks.
\newblock In {\em Proceedings of the IEEE conference on computer vision and
  pattern recognition}, pages 1125--1134, 2017.

\bibitem{8698753}
Yi~Jin, Changjiang Zhou, Xiaodong Teng, Jiatong Ji, Hong Wu, and Jun Liao.
\newblock Pai-wsit: An ai service platform with support for storing and sharing
  whole-slide images with metadata and annotations.
\newblock {\em IEEE Access}, 7:54780--54786, 2019.

\bibitem{kiupel2011proposal}
M~Kiupel, JD~Webster, KL~Bailey, S~Best, J~DeLay, CJ~Detrisac, SD~Fitzgerald,
  D~Gamble, PE~Ginn, MH~Goldschmidt, et~al.
\newblock Proposal of a 2-tier histologic grading system for canine cutaneous
  mast cell tumors to more accurately predict biological behavior.
\newblock {\em Veterinary pathology}, 48(1):147--155, 2011.

\bibitem{maddox2019questions}
Thomas~M Maddox, John~S Rumsfeld, and Philip~RO Payne.
\newblock Questions for artificial intelligence in health care.
\newblock {\em Jama}, 321(1):31--32, 2019.

\bibitem{mcsporran2009histologic}
KD~McSporran.
\newblock Histologic grade predicts recurrence for marginally excised canine
  subcutaneous soft tissue sarcomas.
\newblock {\em Veterinary Pathology}, 46(5):928--933, 2009.

\bibitem{metter2019trends}
David~M Metter, Terence~J Colgan, Stanley~T Leung, Charles~F Timmons, and
  Jason~Y Park.
\newblock Trends in the us and canadian pathologist workforces from 2007 to
  2017.
\newblock {\em JAMA network open}, 2(5):e194337--e194337, 2019.

\bibitem{meuten2016mitotic}
DJ~Meuten, FM~Moore, and JW~George.
\newblock Mitotic count and the field of view area: time to standardize, 2016.

\bibitem{mills2015prognostic}
SW~Mills, KM~Musil, JL~Davies, S~Hendrick, C~Duncan, ML~Jackson, B~Kidney,
  H~Philibert, BK~Wobeser, and E~Simko.
\newblock Prognostic value of histologic grading for feline mammary carcinoma:
  a retrospective survival analysis.
\newblock {\em Veterinary pathology}, 52(2):238--249, 2015.

\bibitem{pantanowitz2020accuracy}
Liron Pantanowitz, Douglas Hartman, Yan Qi, Eun~Yoon Cho, Beomseok Suh,
  Kyunghyun Paeng, Rajiv Dhir, Pamela Michelow, Scott Hazelhurst, Sang~Yong
  Song, et~al.
\newblock Accuracy and efficiency of an artificial intelligence tool when
  counting breast mitoses.
\newblock {\em Diagnostic pathology}, 15(1):1--10, 2020.

\bibitem{paszke2019pytorch}
Adam Paszke, Sam Gross, Francisco Massa, Adam Lerer, James Bradbury, Gregory
  Chanan, Trevor Killeen, Zeming Lin, Natalia Gimelshein, Luca Antiga, et~al.
\newblock Pytorch: An imperative style, high-performance deep learning library.
\newblock {\em Advances in neural information processing systems},
  32:8026--8037, 2019.

\bibitem{pedersen2020fastpathology}
André Pedersen, Marit Valla, Anna~M. Bofin, Javier~Pérez de~Frutos, Ingerid
  Reinertsen, and Erik Smistad.
\newblock Fastpathology: An open-source platform for deep learning-based
  research and decision support in digital pathology, 2020.

\bibitem{pena2013prognostic}
L~Pe{\~n}a, PJ~De Andr{\'e}s, M~Clemente, P~Cuesta, and MD~Perez-Alenza.
\newblock Prognostic value of histological grading in noninflammatory canine
  mammary carcinomas in a prospective study with two-year follow-up:
  relationship with clinical and histological characteristics.
\newblock {\em Veterinary Pathology}, 50(1):94--105, 2013.

\bibitem{pierre2002peripheral}
Robert~V Pierre.
\newblock Peripheral blood film review: the demise of the eyecount leukocyte
  differential.
\newblock {\em Clinics in laboratory medicine}, 22(1):279--297, 2002.

\bibitem{puri2019automated}
Munish Puri, Shelley~B Hoover, Stephen~M Hewitt, Bih-Rong Wei, Hibret~Amare
  Adissu, Charles~HC Halsey, Jessica Beck, Charles Bradley, Sarah~D Cramer,
  Amy~C Durham, et~al.
\newblock Automated computational detection, quantitation, and mapping of
  mitosis in whole-slide images for clinically actionable surgical pathology
  decision support.
\newblock {\em Journal of pathology informatics}, 10, 2019.

\bibitem{rao2018mitos}
Siddhant Rao.
\newblock Mitos-rcnn: A novel approach to mitotic figure detection in breast
  cancer histopathology images using region based convolutional neural
  networks.
\newblock {\em arXiv preprint arXiv:1807.01788}, 2018.

\bibitem{ren2015faster}
Shaoqing Ren, Kaiming He, Ross Girshick, and Jian Sun.
\newblock Faster r-cnn: Towards real-time object detection with region proposal
  networks.
\newblock {\em arXiv preprint arXiv:1506.01497}, 2015.

\bibitem{ronneberger2015u}
Olaf Ronneberger, Philipp Fischer, and Thomas Brox.
\newblock U-net: Convolutional networks for biomedical image segmentation.
\newblock In {\em International Conference on Medical image computing and
  computer-assisted intervention}, pages 234--241. Springer, 2015.

\bibitem{spangler2006histologic}
WL~Spangler and Philip~H Kass.
\newblock The histologic and epidemiologic bases for prognostic considerations
  in canine melanocytic neoplasia.
\newblock {\em Veterinary Pathology}, 43(2):136--149, 2006.

\bibitem{sudre2017generalised}
Carole~H Sudre, Wenqi Li, Tom Vercauteren, Sebastien Ourselin, and M~Jorge
  Cardoso.
\newblock Generalised dice overlap as a deep learning loss function for highly
  unbalanced segmentations.
\newblock In {\em Deep learning in medical image analysis and multimodal
  learning for clinical decision support}, pages 240--248. Springer, 2017.

\bibitem{tan2019efficientnet}
Mingxing Tan and Quoc~V Le.
\newblock Efficientnet: Rethinking model scaling for convolutional neural
  networks.
\newblock {\em arXiv preprint arXiv:1905.11946}, 2019.

\bibitem{veta2016mitosis}
Mitko Veta, Paul~J Van~Diest, Mehdi Jiwa, Shaimaa Al-Janabi, and Josien~PW
  Pluim.
\newblock Mitosis counting in breast cancer: Object-level interobserver
  agreement and comparison to an automatic method.
\newblock {\em PloS one}, 11(8):e0161286, 2016.

\bibitem{voss2015mitotic}
Sarah~M Voss, Meghan~P Riley, Parvez~M Lokhandwala, Ming Wang, and Zhaohai
  Yang.
\newblock Mitotic count by phosphohistone h3 immunohistochemical staining
  predicts survival and improves interobserver reproducibility in
  well-differentiated neuroendocrine tumors of the pancreas.
\newblock {\em The American journal of surgical pathology}, 39(1):13--24, 2015.

\bibitem{Xie2016}
Saining Xie, Ross Girshick, Piotr Dollár, Zhuowen Tu, and Kaiming He.
\newblock Aggregated residual transformations for deep neural networks.
\newblock {\em arXiv preprint arXiv:1611.05431}, 2016.

\end{thebibliography}

\appendix
\section{Supplemental Data}

\begin{table}[h]
\begin{tabular}{l|l|l|l|l|l|}
\cline{2-6}
\multicolumn{1}{c|}{\textbf{}}                                                     & \multicolumn{2}{c|}{\textbf{Algorithm 1, brute-force}}                                                                   & \multicolumn{2}{c|}{\textbf{Algorithm 2, \textit{k}-d tree}}                                                                      & \multicolumn{1}{c|}{\textbf{Comparison}}                         \\ \hline
\multicolumn{1}{|l|}{\begin{tabular}[c]{@{}l@{}}WSI MF\\ Total Count\end{tabular}} & \begin{tabular}[c]{@{}l@{}}MF maxcount \\ in 10 HPF\end{tabular} & \begin{tabular}[c]{@{}l@{}}time \\ (sec)\end{tabular} & \begin{tabular}[c]{@{}l@{}}MF maxcount \\ in 10 HPF\end{tabular} & \begin{tabular}[c]{@{}l@{}}time \\ (sec)\end{tabular} & \begin{tabular}[c]{@{}l@{}}net time saving \\ (sec)\end{tabular} \\ \hline
\multicolumn{1}{|l|}{1}                                                            & 1                                                                & 0.01                                                  & 1                                                                & 0.00                                                  & 0.01                                                             \\ \hline
\multicolumn{1}{|l|}{240}                                                          & 19                                                               & 2.48                                                  & 33                                                               & 0.02                                                  & 2.45                                                             \\ \hline
\multicolumn{1}{|l|}{500}                                                          & 55                                                               & 14.92                                                 & 94                                                               & 0.21                                                  & 14.70                                                            \\ \hline
\multicolumn{1}{|l|}{750}                                                          & 103                                                              & 41.81                                                 & 200                                                              & 0.69                                                  & 41.12                                                            \\ \hline
\multicolumn{1}{|l|}{959}                                                          & 148                                                              & 93.72                                                 & 103                                                              & 1.57                                                  & 92.15                                                            \\ \hline
\end{tabular}
\caption{Comparing 10HPF search algorithms, the effect of proposed \textit{k}-d tree based search versus brute-force based search.}
\label{tab:tenHPF_time_comparison}
\end{table}

\begin{figure}[h]
\caption{Comparing mitotic counts of canine (C) and feline (F) neoplasia determined by pathologists without AI assistance (Non-AI assisted MC), by AI alone (AI-only MC), and by pathologists with AI assistance (AI-Assisted MC), and the resulting changes in grading when applicable. "Up" and "Down" arrows indicate whether the AI-assisted MC average increased or decreased as compared to the Non-AI asissted MC by pathologist.}
\label{fig:Mitotic_index_determination_with_and_without_AI_assistance}
\includegraphics[scale=0.80]{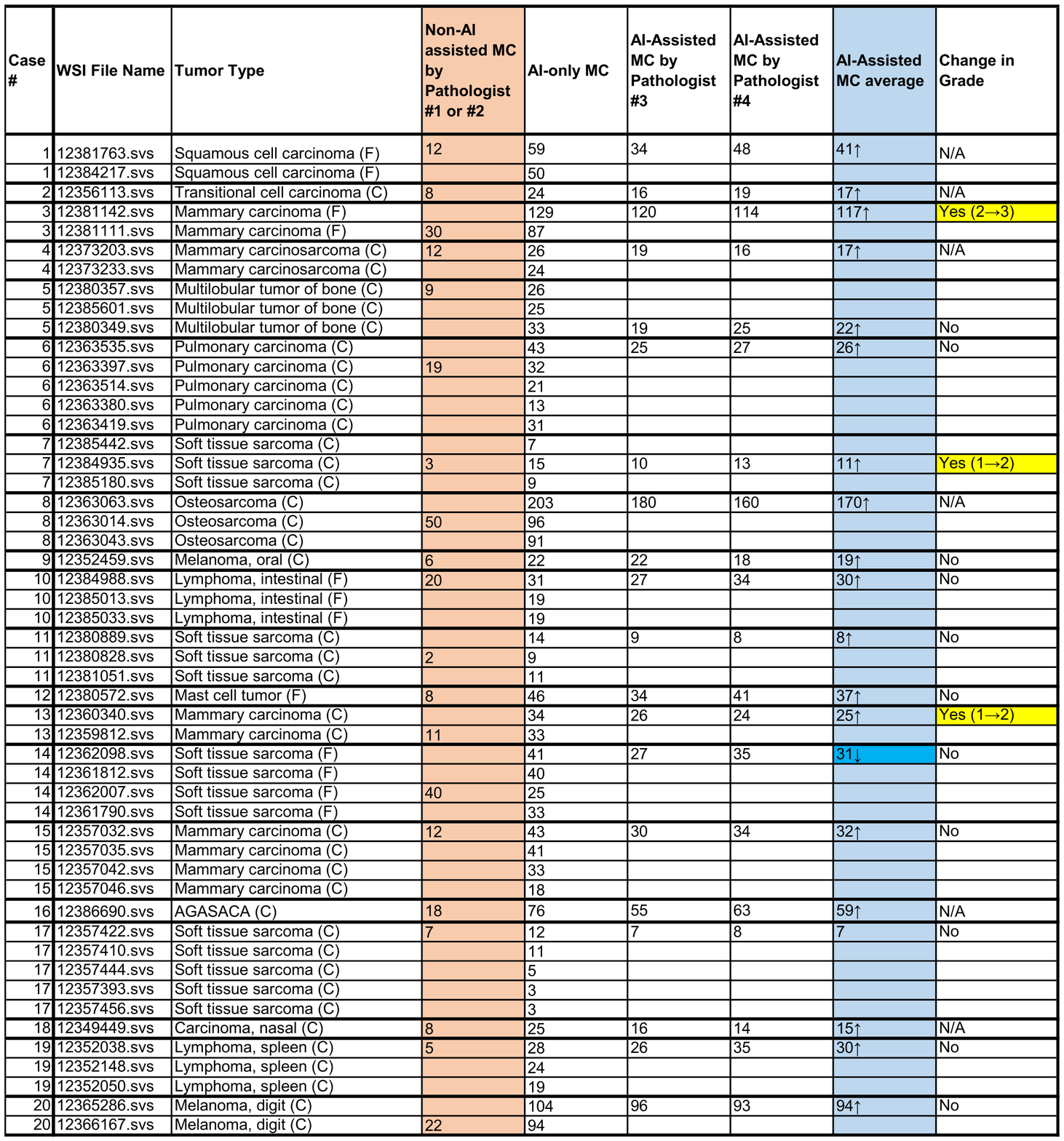}
\centering
\end{figure}

\begin{figure}[h]
\includegraphics[scale=0.80]{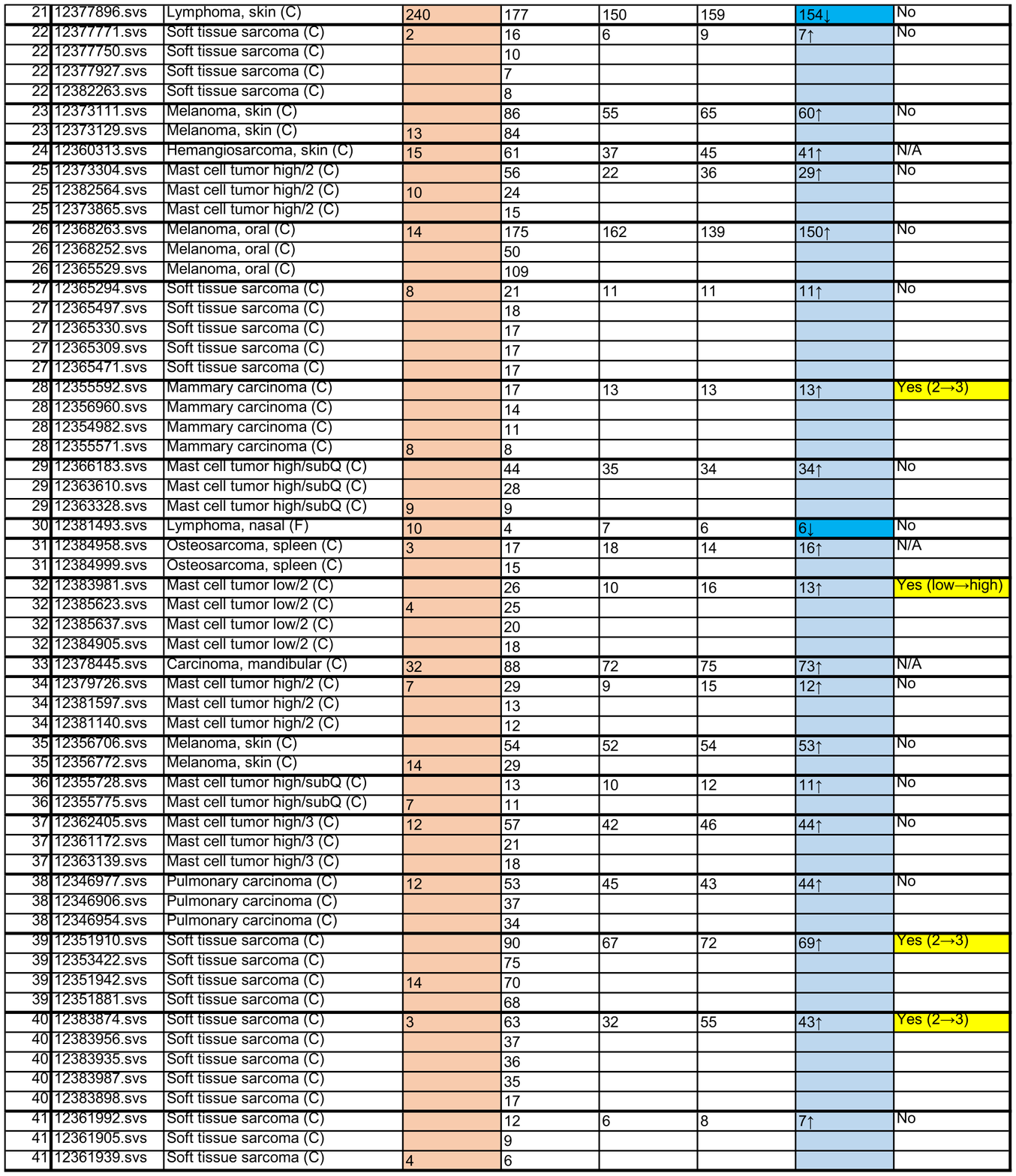}
\centering
\end{figure}

\begin{figure}[h]
\includegraphics[scale=0.75]{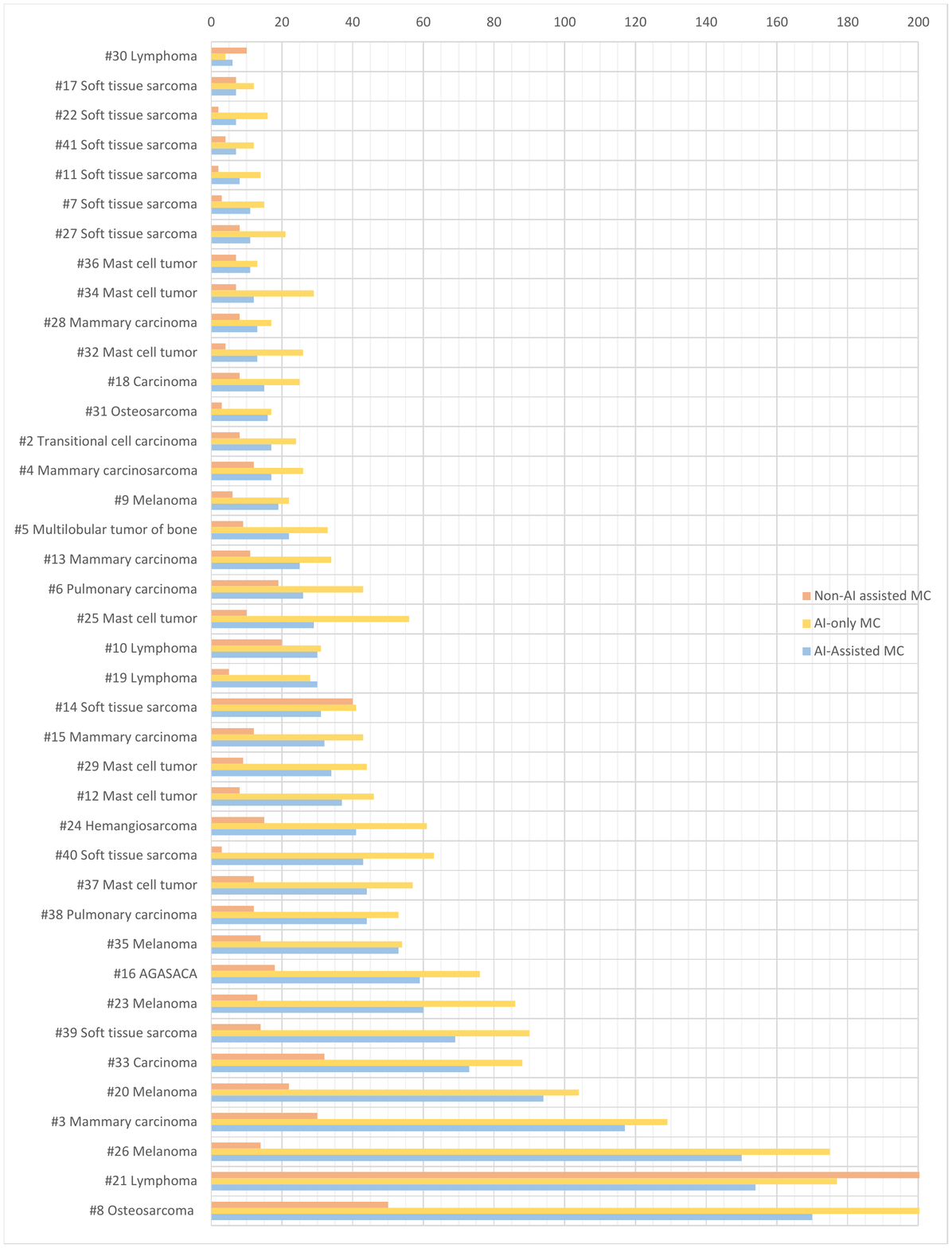}
\centering
\end{figure}

\end{document}